# ComVi: Context-Aware Optimized Comment Display in Video Playback

Minsun Kim
KAIST
Daejeon, Republic of Korea
sunnykimhappy@kaist.ac.kr

Dawon Lee*
Kookmin University
Seoul, Republic of Korea
dawon.lee@kookmin.ac.kr

Junyong Noh*
KAIST
Daejeon, Republic of Korea
junyongnoh@kaist.ac.kr

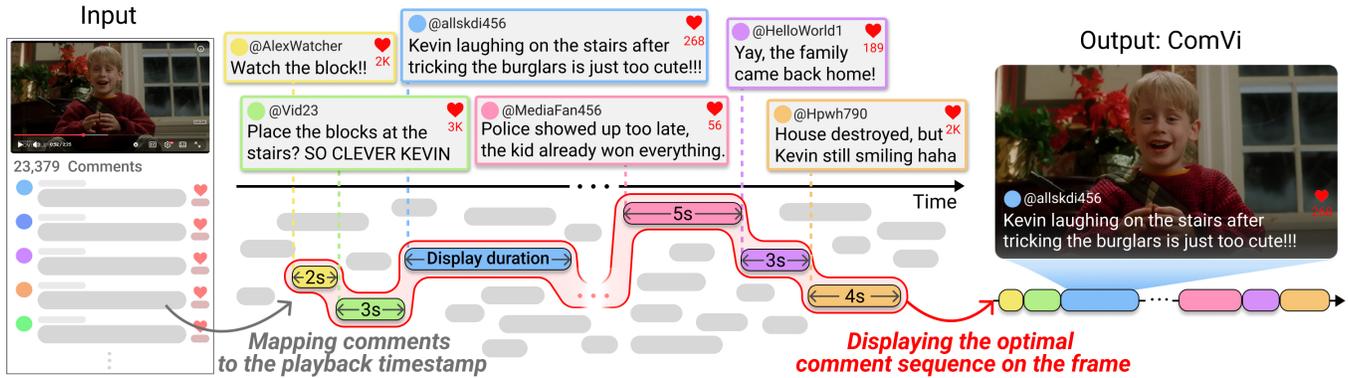

Figure 1: Overview of *ComVi*. Given a video and its comments (*Input*), ComVi first maps each comment to semantically relevant timestamps by computing audio-visual correlations. It then selects an optimal comment sequence by balancing temporal semantic relevance, popularity, and adequate display durations. Finally, the selected comments are presented on the video frame at their corresponding timestamps during playback (*Output*). Video source: V3 in Table A.1.

## Abstract

On general video-sharing platforms like YouTube, comments are displayed independently of video playback. As viewers often read comments while watching a video, they may encounter ones referring to moments unrelated to the current scene, which can reveal spoilers and disrupt immersion. To address this problem, we present *ComVi*, a novel system that displays comments at contextually relevant moments, enabling viewers to see time-synchronized comments and video content together. We first map all comments to relevant video timestamps by computing audio-visual correlation, then construct the comment sequence through an optimization that considers temporal relevance, popularity (number of likes), and display duration for comfortable reading. In a user study, ComVi provided a significantly more engaging experience than conventional video interfaces (i.e., YouTube and Danmaku), with 71.9% of participants selecting ComVi as their most preferred interface.

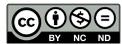

## ACM Reference Format:

## 1 Introduction

Video-sharing platforms like *YouTube*, *TikTok*, and *Vimeo* host billions of daily comments that provide viewers with opportunities to gain deeper insights and enjoy richer experiences with video content [25, 34]. These comments often appear alongside the content, offering background information [34, 44], contextual cues [44], and highlighting key moments [58] or overlooked details [44]. They also expose users to diverse interpretations that broaden perspectives [34, 46] and foster a sense of shared experience among viewers [46]. Through these functions, comments enhance the way viewers interpret and relate to video content, adding layers of insight beyond the video itself [46].

As part of this enhanced viewing experience, many people also read comments *simultaneously during video playback* [15, 31, 47]. To support this behavior, some platforms provide Danmaku-based interfaces (e.g., *NicoNico*, *Bilibili*, and *ViKi*) that enable viewers to see comments synchronized with video playback [13, 14, 21, 39, 78] (Figure 2). However, Danmaku-based interfaces require each comment to have a predefined display timestamp [13, 14, 78], making them inapplicable to general comments that do not have such temporal metadata. Moreover, viewers may miss content when multiple comments are displayed at the same time, as there is often insufficient time to read each comment before it disappears from the screen [88].

When Danmaku-based interfaces are not utilized, comments are typically displayed without synchronization to video playback on the general video-sharing platforms such as *YouTube*, *TikTok*, and *Vimeo*; the comments remain static and independent of the video timeline. As a result, viewers may encounter comments that reference unrelated moments or scenes, appearing either too early or too late relative to the current context. For example, viewers may see a comment such as *"Plot twist at the end—the cop was a double agent!"* early in the video, leading to spoilers of plot developments, twists, or endings, fundamentally diminishing their viewing enjoyment and engagement. Although prior research has explored selectively presenting highly relevant comments by analyzing their relation to the overall video content [15, 24], such approaches do not address the problem of temporal alignment as the selected comments are presented without reference to a specific timestamp.

In response, we introduce *ComVi*, a novel system that displays comments on general video platforms at temporally appropriate moments (or timestamps) aligned with the currently viewed scene, enabling viewers to enjoy video content together with the relevant comments (Figure 1). To achieve this, ComVi (1) displays contextually relevant and popular comments drawn from the vast pool of comments synchronized with the video timeline, (2) ensures adequate display duration for viewers to fully read each comment. In the first step, each comment is aligned with multiple relevant video timestamps by computing the audio-visual semantic correlation with the corresponding scene. In the second step, these aligned comments are assigned with display durations estimated based on their text lengths to generate all possible non-overlapping comment sequences. Finally, the optimal sequence is obtained through dynamic programming that jointly considers the correlation and number of likes. We further present user-driven comment curation customization to support individual reading behaviors.

The user study ($N = 32$) showed that ComVi achieved a significantly more engaging comment-integrated viewing experience than conventional video interfaces, with lower physical demand than *YouTube* and lower mental demand than *Danmaku*. This was further reflected in the preference survey, where 71.9% of participants selected ComVi as their most preferred interface. Altogether, our contributions can be summarized as follows:

- We present a novel approach that displays comments at semantically relevant timestamps in the video playback timeline.
- We propose a method that selects the optimal comment sequence by balancing semantic relevance, popularity, and adequate display duration.
- We introduce a new way that provides viewers with an enhanced video comment viewing experience compared to conventional platforms, as demonstrated by the user study.

## 2 Related Work
## 2.1 Supporting Comment Reading in Various Video-Watching Environments

Users often read comments during video playback in various video-watching environments. These include reading comments posted by other viewers on online video-sharing platforms [15, 31, 47], following real-time comments on live streaming services [45], and engaging in second-screen activities, where viewers watch TV broadcasts while simultaneously following comments on social media platforms on another screen [7, 77].

On general video-sharing platforms, a single video may accumulate thousands of comments, which can easily overwhelm users [15]. To address this challenge, researchers have developed methods that either detect irrelevant comments (e.g., spam, advertisements) [2, 24, 51, 79] to filter them out from the large pool of comments, or detect relevant comments [15, 24] to present them as more useful ones for viewers. Similar to these approaches, ComVi also presents a subset of relevant ones selected from the large pool of comments. The difference is that it further synchronizes these comments with temporally relevant content during playback.

In Danmaku-based interfaces, a large number of comments often simultaneously clutter the screen, obscuring the underlying video content [43, 81] (see Figure 2A, B). To mitigate this, some systems consolidate identical or semantically similar comments into a single one and adjust the font size proportionally to their frequency [13, 64, 88], thereby reducing screen clutter while emphasizing frequently mentioned topics. Another approach involves categorizing comments by content or sentiment and allowing users to filter the displayed comments based on their preferred categories [8, 88].

In live streaming environments with large audiences, real-time comments arrive so rapidly that viewers often find it difficult to follow and comprehend them [26, 29, 45, 52]. To address this, some systems show only those comments that meet a minimum vote threshold from other users [40, 45, 67]. Aghahoseini et al. [1] visualized frequently used words and emojis in real-time comments with sizes proportional to their appearance frequency, helping viewers to quickly grasp the content of actively discussed topics at a glance without having to read every individual comment.

For TV viewing, the most common way to support reading comments from social media platforms is to overlay them on the TV screen [7], allowing viewers to follow comments without needing to look at their mobile devices. For example, Centieiro et al. [10] proposed a system that displays comments or emojis that users type on their mobile phones on the TV screen. These approaches across diverse video-watching environments—such as *Danmaku*, live streaming, and TV—operate on comments whose display times are predefined by the posting moment. In contrast, our work focuses on general comments that lack such temporal metadata.

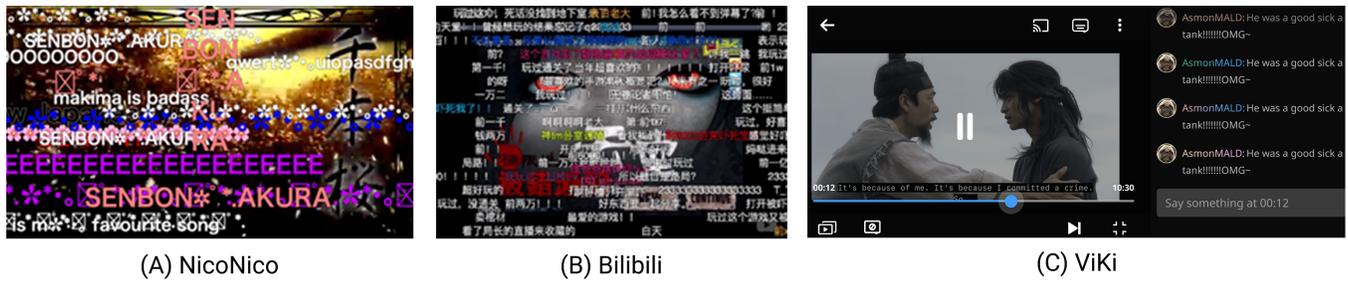

(A) NicoNico	(B) Bilibili	(C) ViKi

Figure 2: Danmaku-based interfaces across various platforms. In such interfaces, when commenters post comments at specific timestamps while watching a video, those timestamps are stored as the display moments for the respective comments. When other viewers play the video, the comments appear at their assigned timestamps, either flowing horizontally from right to left across the video screen (A, B) or displayed vertically in a separate side panel (C).

## 2.2 Synchronizing Textual Content with Specific Video Playback Timestamps

Researchers have developed interfaces that synchronize textual content with specific video timestamps to support navigation and enhance comprehension of video content. To improve video navigation, some systems synchronize transcripts with playback timestamps and present them in a side or bottom panel, enabling users to jump directly to specific moments by clicking the corresponding text segment [12, 27, 48, 53–55, 60, 73]. These alignments were typically achieved either through manual human annotation or by leveraging automatic speech-to-text models [18]. In the context of music videos, Ma et al. [42] proposed an automated pipeline that synchronizes lyrics with video playback. The resulting video presents each word with animation effects, helping viewers to read the lyrics in sync with the song. The start and end times of each word are determined by a lyric alignment deep learning model [19].

Beyond transcripts, researchers have proposed methods that align various textual sources with video, including educational concepts [38], Q&A comments [83], academic papers [30], plot summaries [53], plot synopses [68, 69], and books [70, 90]. For example, to support learning in educational contexts, Liu et al. [38] generated a concept map—a graph-like visualization showing key concepts and their relationships—linked to the video timestamps where those concepts are explained. The linked concepts were displayed in a side panel, enabling learners to click on a concept and immediately jump to the corresponding part of the video for further explanation. Similarly, Yarmand et al. [83] introduced a method that links user-generated Q&A comments to the timestamps at which the relevant concepts are discussed, showing them in a side panel as playback reaches those moments so that learners can immediately resolve their questions. In both systems [38, 83], human annotators manually link textual content to contextually relevant video timestamps.

Kim et al. [30] aligned passages from research papers with the moments that explained those parts in the presentation videos, displaying the corresponding video segments alongside the paper passages to support deeper comprehension. Pavel et al. [53] temporally aligned plot summaries, scripts, and movie captions with corresponding video timestamps, displaying them side by side. This alignment allows users to jump to specific scenes by searching through any of these textual materials–whether looking for a plot point in the summary, a character name in the script, or specific dialogue in the captions. These approaches [30, 53] align text with video by using Natural Language Processing techniques [62, 76] to measure the similarity between the text and the video's content. In this paper, we extend this line of work to comments from general video-sharing platforms, exploring how to align comments with contextually relevant video timestamps to enable synchronized comment viewing during playback.

## 3 ComVi: The Method

ComVi takes as input a video, its associated set of comments, and the metadata of each comment (i.e., number of likes, user profile)[1]. It then produces a single optimal comment sequence, with the appearance timing and duration of each comment. In the resulting sequence, comments do not overlap in time; one comment is displayed at any moment (timestamp) during video playback.

We begin by converting all comments–those typically found on standard video platforms–into *timed comments* by computing audio-visual correlation (Section 3.1). From these timed comments, we generate all possible sequences by selecting comments while considering their individual reading times (Section 3.2). To evaluate the quality of each sequence, we assign a score to each timed comment within it based on the correlation and number of likes (Section 3.3). We then use dynamic programming [6, 22] to obtain the optimal sequence—the one with the highest total score (Section 3.4). Building on the pipeline, we present extensions of ComVi to support personalized comment curation (Section 3.5). In the following, we detail each process.

### 3.1 Mapping General Comments into Timed Comments

We represent the set of comments $\{C_1, C_2, \cdots, C_I\}$, where $I$ denotes the total number of comments. When multiple comments share identical text, only the one with the most likes is retained to avoid duplication. We map each comment $C_i$ into a set of *timed comments* $C_{i,t}$, where each $C_{i,t}$ represents comment $C_i$ scheduled to appear at timestamp $t \in [1, T]$, and $T$ is the video length in seconds. A single comment can be mapped to one or more timestamps in the video where the audio-visual correlation exceeds a predefined threshold (empirically set to 0.3). The correlation is calculated as follows:

$$Corr(C_{i,t}) = \text{Norm}\left(\sqrt{\frac{(Corr_A(C_{i,t}))^2 + (Corr_V(C_{i,t}))^2}{2}}\right), \quad (1)$$

where $Corr_A(C_{i,t})$ and $Corr_V(C_{i,t})$ denote the semantic correlation between comment $C_i$ and the audio and visual content at timestamp $t$, respectively.

---
[1]We collect videos from *YouTube*, along with all the comments and their metadata via crawling.

We compute these correlations using cosine similarity between sentence embeddings, where all textual elements are encoded with Sentence-BERT[2] [57]. For $Corr_A(C_{i,t})$, we use the subtitle text[3] aligned with the video timestamp as the audio representation. The subtitle segment that contains timestamp $t$ is identified for the computation of the correlation between comment $C_i$ and that subtitle segment's text. For $Corr_V(C_{i,t})$, we first segment the video into shots using PySceneDetect [9], then generate a textual description for each shot using the Tarsier video captioning model [85][4]. The video shot that contains timestamp $t$ is identified for the computation of the correlation between comment $C_i$ and that shot's description. The combined correlation scores are normalized to a range between 0 and 1 using the Norm(·) function. For comments containing explicit timestamp references written by the commenter (e.g., *"The goal at* `30:53` *was offside but not called."*), we directly map the comment to the mentioned timestamp(s) to reflect the human commenter's intention.

### 3.2 Generating Candidate Sequences

We generate all candidate sequences[5] from the pool of timed comments that satisfy the following two rules. First, each selected timed comment $C_{i,t}$ must be displayed long enough to allow viewers to fully read it. Specifically, if comment $C_{i,t}$ requires a reading time $Reading(C_i)$, the next comment $C_{i',t'}$ must appear at timestamp $t'$ that satisfies $t' \geq t + Reading(C_i)$. The reading time $Reading(C_i)$ (in seconds) is calculated based on the comment length [17, 65] $L(C_i)$, measured as the number of characters as follows:

$$Reading(C_i) = \min(\alpha_{user} \cdot L(C_i), \tau_{\max}),$$

where $\alpha_{user}$ denotes the average reading speed per character, set to 0.068 seconds by default following Trauzettel-Klosinski et al. [72]. To accommodate individual differences in reading speed [17, 66, 72], viewers may adjust $\alpha_{user}$ according to their preferences. We also impose a maximum display duration $\tau_{\max}$ of 6 seconds, as longer display durations have been shown to negatively influence viewer enjoyment [65]. This limit can also be customized by the viewer. Since our method operates with a temporal resolution of one second, any fractional reading times are rounded up to the nearest whole second. This ensures a minimum display duration of one second per comment, following widely adopted subtitle guidelines that recommend at least one second of on-screen time [4, 50]. As a second rule, to prevent repetitive viewing, once a comment has been selected in the sequence, it is excluded from subsequent selections.

### 3.3 Evaluating Sequence Quality

To evaluate the quality of a sequence, we assign a score $Score(C_{i,t})$ to each timed comment within it. Higher scores are given to comments that are both semantically relevant to the video content and popular among viewers, as measured by the number of likes [41, 61, 63]:

$$Score(C_{i,t}) = \left(w_{corr}Corr(C_{i,t}) + w_{likes}Likes(C_i)\right) \cdot Reading(C_i).$$

Here, $Likes(C_i)$ denotes the normalized like count of a comment $C_i$. The weights were empirically set to $w_{corr} = 2$ and $w_{likes} = 1$. $w_{likes}$ is a user-adjustable parameter, allowing viewers to reduce the influence of popularity by decreasing its value (down to 0) according to their individual preference. We also include the reading time $Reading(C_i)$ as a weighting factor to mitigate selection bias toward short comments. Without this factor, shorter comments may be selected more frequently within a fixed video duration, leading to inflated total scores. By weighting with reading time, we reduce this bias and promote a more balanced selection.

Because raw like counts in a single video often follow a long-tailed distribution (i.e., most comments have few likes while a few have extremely high ones) [3, 41], we normalize them using a Box-Cox transformation following Luo et al. [41]. Specifically, the normalized like count is computed as:

$$Likes(C_i) = \begin{cases} \text{Norm}\left(\text{BoxCox}(l_i)\right) & \text{if } l_i \neq 0 \\ 0 & \text{otherwise} \end{cases},$$

where $l_i$ is the raw like count of comment $C_i$. Because the Box–Cox transformation operates only on positive values, we apply the transformation to comments with non-zero like count; otherwise, we assign a zero value. The BoxCox(·) is defined as:

$$\text{BoxCox}(l_i) = \begin{cases} \dfrac{l_i^\lambda - 1}{\lambda} & \lambda \neq 0 \\ \ln(l_i) & \lambda = 0 \end{cases},$$

where $\lambda$ is a transformation parameter automatically estimated from the like count distribution to transform the values so that they approximately follow a normal distribution[6]. After applying this, we normalize the transformed values to a range between 0 and 1 using the Norm(·) function.

### 3.4 Selecting the Optimal Sequence

Given the candidate comment sequences $\mathcal{S}$, the optimal comment sequence $\mathcal{S}^*$ is obtained by maximizing the summation of individual timed comment scores in it, as defined by:

$$\mathcal{S}^* = \arg\max_{\mathcal{S}} \sum_{k=1}^{n} Score(C_{i_k, t_k}),$$

where each $C_{i_k, t_k}$ denotes the $k$-th selected timed comment in the sequence. The total number of selected comments $n$ is automatically determined during the candidate sequence generation process (Section 3.2), based on the video length and the reading time rule.

---

[2]We used the pretrained `all-mpnet-base-v2` model.
[3]If the content does not provide subtitle text, it can be generated using a speech-to-text model such as Whisper [56].
[4]Prompted with *"Describe the video in detail"*.
[5]One example could be $C_{7,5} \to C_{1,10} \to C_{25,17} \to \cdots$. Here, comments $C_7$, $C_1$, and $C_{25}$ are displayed at 5 second, 10 second, and 17 second on the screen, respectively.

[6]We used `scipy.stats.boxcox` to automatically estimate the optimal $\lambda$ parameter.

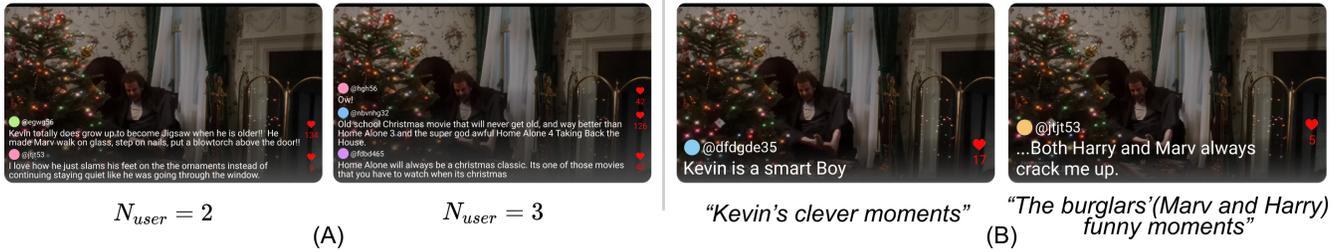

Figure 3: Display changes of ComVi after applying personalized comment curation features: (A) controlling the maximum number of concurrently displayed comments ($N_{user}$ = 2 and 3), and (B) applying the user-specified query $q_{user}$ (queries shown below each image). All usernames and profile images were replaced to protect personal information. Video source: V3 in Table A.1.

## 3.5 ComVi Extensions: Supporting Personalized Comment Curation

On top of the automated comment curation, ComVi provides user-driven customization to support individual reading behaviors. Beyond the single-comment display, ComVi allows viewers to adjust the maximum number of comments to be displayed at once ($N_{user}$), supporting skim reading or vibe-catching by focusing on keywords across multiple comments (Figure 3A). This function is implemented by only modifying the first rule of selecting the next timed comment ($t' \geq t + Reading(C_i)$) in candidate sequence generation (Section 3.2) to $t' \geq t$ while ensuring that $Overlap(C_{i',t'}) < N_{user}$. $Overlap(C_{i',t'})$ is the number of previously selected timed comments whose end display time (i.e., start time + reading time) is later than $t'$, defined as:

$$Overlap(C_{i',t'}) = \left|\left\{C_{i,t} \in \mathcal{S}_{current} \mid t + Reading(C_i) > t'\right\}\right|,$$

where $\mathcal{S}_{current}$ denotes a set of timed comments selected so far during the candidate sequence generation process. This ensures that no more than $N_{user}$ comments are simultaneously visible at once while preserving the full reading time of each selected comment[7].

Another extension is to generate a sequence consisting of comments relevant to viewers' interests (Figure 3B). When a user inputs a natural language query $q_{user}$, comments are filtered from the entire set as a preprocessing step before executing the process described in Section 3.1. We first compute the cosine similarity between the embeddings[8] of $q_{user}$ and each comment. Only comments with similarity scores above a predefined threshold (empirically set to 0.6) then proceed to the next mapping process.

## 4 Results

Figure 4 illustrates the structure of the resulting comment sequence generated by ComVi and how it is displayed on a video frame. In the figure example, we used a 3-minute and 34-second video (V1 in Table A.1) containing 14,880 comments, with the number of likes ranging from 0 to 21,000 ($M = 20.46, SD = 340.18$). Through our computation, 46 comments were selected for the resulting sequence. The average display duration of the comments was 4.65 seconds (SD = 1.75). The average number of likes for these comments was 1,069.24 (SD = 2249.68).

---

[7]For example, suppose $N_{user} = 2$ and two comments have been selected so far: $C_{5,2}$ with end time 5 and $C_{3,3}$ with end time 7. The next timed comment $C_{37,4}$ would be ineligible because $Overlap(C_{37,4}) = 2$ at timestamp 4.
[8]We use the same Sentence-BERT model for encoding as in Section 3.1.

## 4.1 Results of Semantic Correlation

Figure 5 shows visual examples of ComVi across various video genres, such as documentary, movie, and news. ComVi successfully presented comments that are contextually aligned with the content unfolding on screen. For example, in the documentary, during a scene in which the cat carefully avoided stepping on the leaf and jumped over it, ComVi displayed a commenter's evaluation of that action (*"This cat is so polite, it didn't disturb the leaf and just gently jumped over it."*). In the movie, when the burglar steps on the glass ornaments and reacts in pain, ComVi displayed the commenter's humorous and empathetic reaction to that incident (*"The glass ornaments get me every time. Ouch!"*). In the news, when the anchor provided an explanation that contained an error, ComVi displayed the commenter's critical reaction pointing out the mistake (*"She said bill was $42... gave him a $50 and she said 'there's 7 dollars left' I guess math is not required to do reporting for CNN."*). For more video results, please refer to our project page: https://w-dlee.github.io/comvi.

For further analysis of contextual alignment, we computed the average correlation scores of the comment sequences generated by our method (*ComVi*) and two baseline conditions, *Ground-truth* and *Random*, across three videos. The correlation score of each comment in the sequence was calculated using Equation (1). In the *ComVi* condition, we neutralized the effect of popularity by setting $Likes(C_i) = 0$ in the scoring function to examine how correlation alone contributes to the quality of the comment sequence. In the *Ground-truth* condition, we leveraged comments containing timestamp references (e.g., *"The goal at 30:53 was offside but not called."*). Because these comments were written by users who explicitly referred to specific video moments [84], they can serve as strong evidence of contextual relevance at those timestamps. For each such comment, we computed the correlation at its referenced timestamp(s) and obtained the highest value. We then selected the comments with the highest correlation scores, matching the total number used in *ComVi*. In the *Random* condition, we randomly selected an equal number of comments from the entire pool and placed them at the identical timestamps corresponding to *ComVi*.

*ComVi* substantially outperformed the *Random* condition, while achieving correlation scores comparable to the *Ground-truth*: documentary ($N = 32$; *ComVi* : $M = 0.66, SD = 0.16$; *Ground-truth* : $M = 0.76, SD = 0.02$; *Random* : $M = 0.14, SD = 0.12$), movie ($N = 42$; *ComVi* : $M = 0.66, SD = 0.19$; *Ground-truth* : $M = 0.61, SD =$

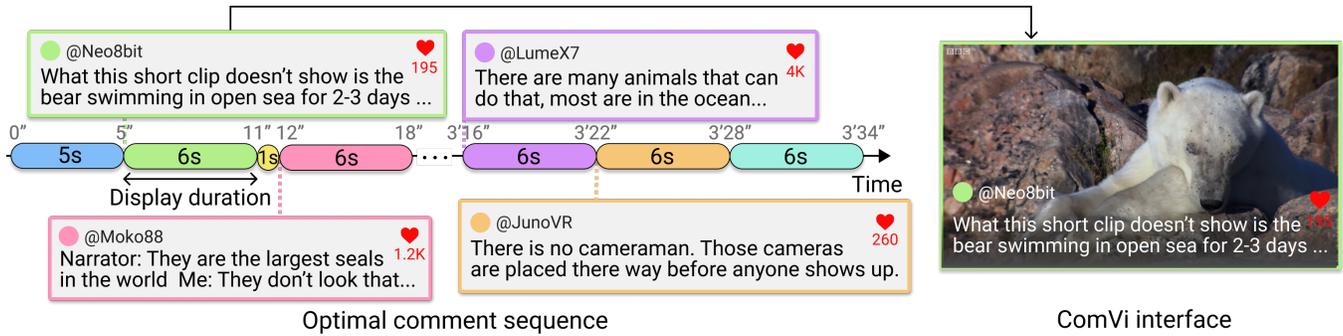

Figure 4: The left panel visualizes how the resulting comment sequence is constructed along the timeline. Each colored bar represents a selected comment, and its starting position on the timeline indicates when it appears in the interface, with its length representing the duration of the display. The right panel illustrates how the resulting comment sequence is presented in the interface. Each comment is displayed at the bottom of the screen, along with its metadata (i.e., user profile and like count). All usernames and profile images were replaced to protect personal information. Video source: V1 in Table A.1.

0.12; $Random: M = 0.17, SD = 0.10$), and news ($N = 34$; $ComVi : M = 0.74, SD = 0.11$; $Ground\text{-}truth : M = 0.72, SD = 0.04$; $Random : M = 0.24, SD = 0.13$). This demonstrates that ComVi successfully presents contextually relevant comments. In some cases, the value of *ComVi* was slightly lower than *Ground-truth*, due to our reading time rule (Section 3.2): when generating the sequence, our method skips a high-correlation timed comment if its display duration overlaps with another comment already included in the sequence. Instead, it selects a different timed comment that satisfies the reading time rule, even if its correlation score is slightly lower. In other cases, the value of *ComVi* was slightly higher than *Ground-truth*, because a comment containing explicit timestamp references may not only refer to the audio-visual content of the scene but also provide background information related to that moment [84] (e.g., a comment explaining the historical context of an event shown on screen), which can result in lower audio-visual correlation.

### 4.2 Results of Popularity

To analyze whether ComVi effectively reflects the popularity, we compared the comment sequences generated by our method (*ComVi*) with two baseline conditions: the entire set of comments (*Total*), and a comment sequence generated without considering the normalized like count term $Likes(C_i)$ (*Likes-ablated*), which is equivalent to the *ComVi* condition described in Section 4.1. Figure 6 presents normalized like count distributions for each condition. As shown in the figure, *ComVi* substantially outperformed both *Likes-ablated* and *Total*, while the mean value in the *Likes-ablated* condition remained very low, showing only a marginal increase over *Total*: documentary ($ComVi : N = 25, M = 0.94, SD = 0.11$; $Likes\text{-}ablated : N = 32, M = 0.14, SD = 0.30$; $Total : N = 66496, M = 0.10, SD = 0.25$), movie ($ComVi : N = 34, M = 0.76, SD = 0.21$; $Likes\text{-}ablated : N = 42, M = 0.31, SD = 0.33$; $Total : N = 1689, M = 0.15, SD = 0.27$), news ($ComVi : N = 29, M = 0.93, SD = 0.08$; $Likes\text{-}ablated : N = 34, M = 0.12, SD = 0.28$; $Total : N = 75358, M = 0.09, SD = 0.23$). These results indicate that ComVi successfully presents comments with high popularity.

### 4.3 Impact of Reading Speed on the Comment Sequence Structure

Figure 7 illustrates how changes in the reading speed parameter $\alpha_{user}$ affect the number of selected comments and the display duration of each comment in the resulting sequence. We used three $\alpha_{user}$ values–0.048, 0.068, and 0.088–chosen to lie within the recommended range for displaying subtitles [32][9]. As $\alpha_{user}$ increased, the display duration of each comment correspondingly increased ($M = 3.45, SD = 1.65$; $M = 3.89, SD = 1.82$; and $M = 4.63, SD = 1.72$, respectively), while the total number of comments included in the sequence decreased (85, 75, and 63 respectively). This inverse relationship naturally occurred because longer display durations per comment reduce the total number of comments that can be accommodated within a fixed video duration.

### 4.4 Results of Customized Comment Curation

Figure 8 illustrates the effect of the maximum number of concurrently displayed comments $N_{user}$ on the sequence composition. As $N_{user}$ increases from 2 to 3, the total number of comments also rose from 65 to 96 (approximately 1.5 times) while successfully ensuring that the number of concurrently displayed comments did not exceed $N_{user}$. Figure 9 shows how applying the user-specified query $q_{user}$ modifies the composition. As shown in the figure, applying $q_{user}$ alters the comment sequence to consist of a different set of comments relevant to the content of $q_{user}$. For video results, please refer to our project page: https://w-dlee.github.io/comvi.

### 4.5 Implementation

ComVi was implemented in *Python* and executed on a machine equipped with an AMD EPYC 7352 2.30GHz 24-Core CPU, 62GB RAM, and a single NVIDIA RTX A5000 GPU. The computation time ranges from 17 seconds to 1 minute 11 seconds across five videos (2 to 10 minutes in length) with over a thousand comments each[10]. This time may vary depending on the number of comments and the video length. It excludes data preparation steps such as gathering videos, comments, and metadata, and obtaining textual representations of visual content required for the subsequent correlation computation. The time required to obtain the textual representations includes segmenting the video into shots (i.e., running `PySceneDetect`), which took around 20 seconds, and generating a textual description for each segmented shot (i.e., running the Tarsier video captioning model), which took around 10 minutes.

---
[9]Measured in characters per second (cps), ranging from 10 cps (0.10 seconds per character) to 20 cps (0.05 seconds per character).

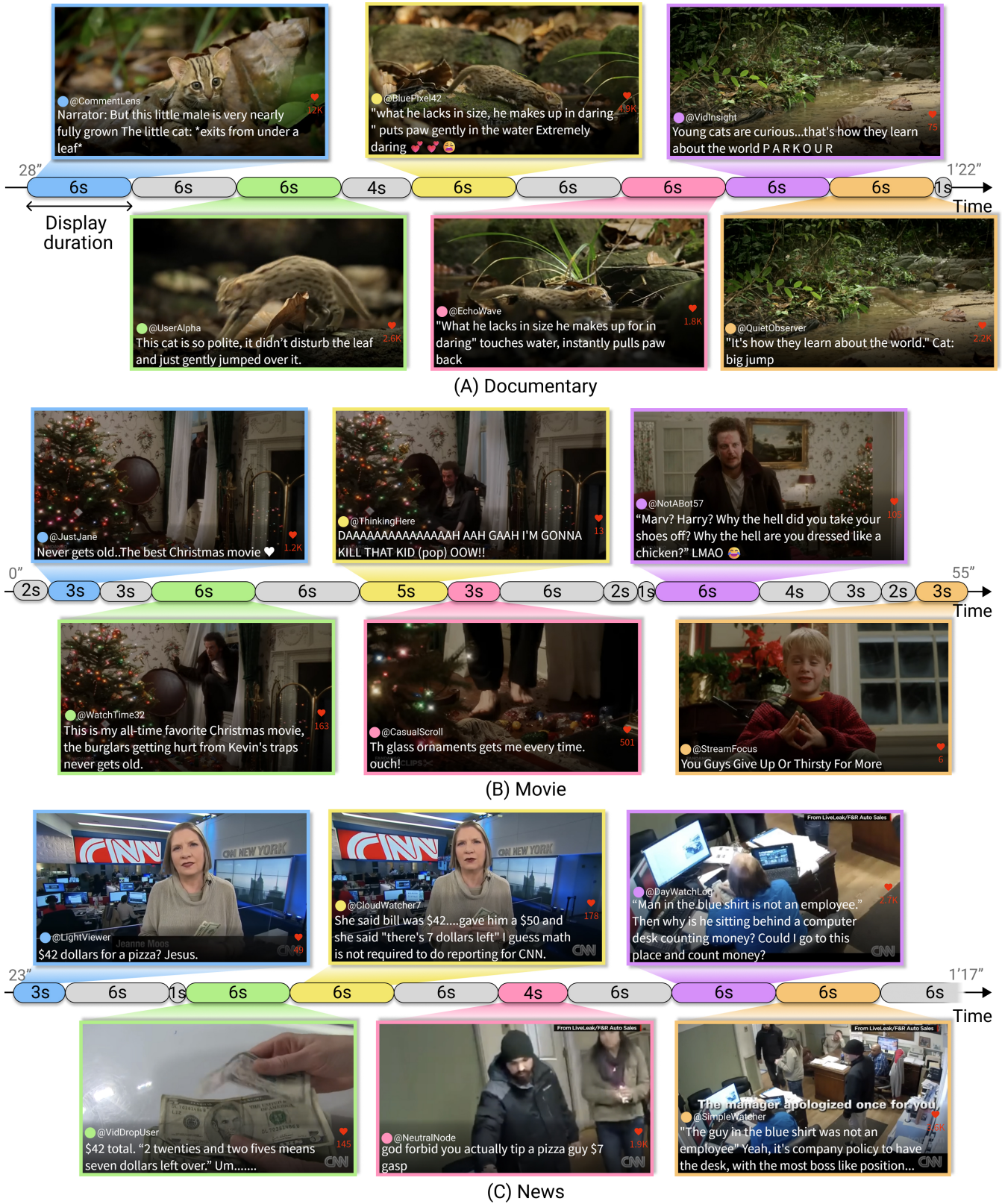

Figure 5: Visual examples of ComVi applied to videos V2 (documentary), V3 (movie), and V4 (news) listed in Table A.1. On the timeline, each bar represents a comment from the optimal sequence, with its starting position indicating the appearance time and its length indicating the display duration. The screenshots aligned above or below each colored bar illustrate how that specific comment is visually presented in ComVi interface. All usernames and profile images were replaced to protect personal information.

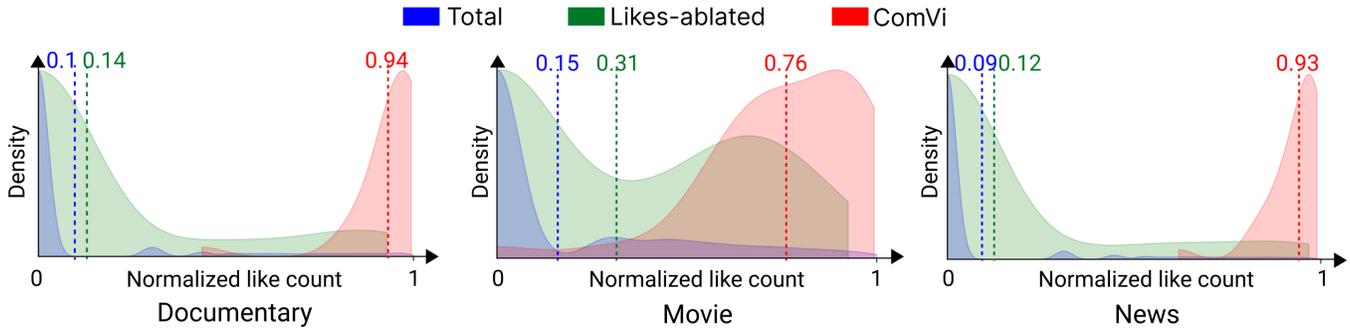

Figure 6: Distributions of normalized like counts under the three conditions (*ComVi*, *Total*, and *Likes-ablated*). Each distribution was visualized using kernel density estimation (KDE), with the y-axis normalized to a range between 0 and 1. The dashed vertical lines indicate the mean value of each condition. The video data used for this analysis is identical to that in Section 4.1.

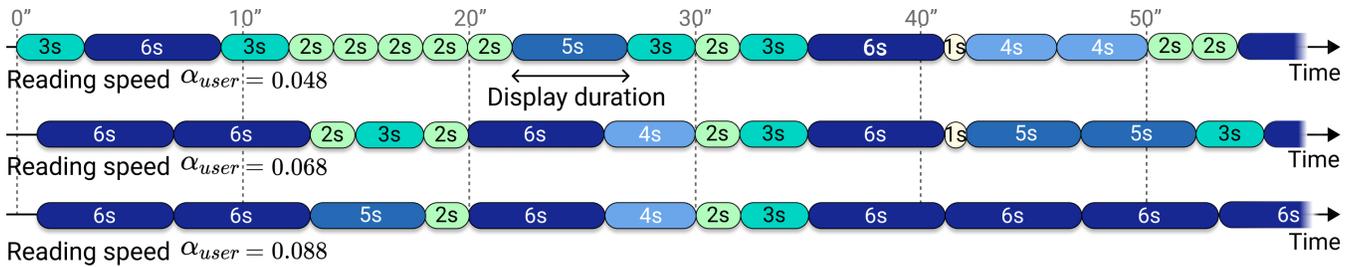

Figure 7: Impact of the reading speed parameter $\alpha_{user}$ on the composition of the resulting sequence. Each bar represents a comment, with its position along the video timeline indicating when it appears and its length corresponding to the comment's display duration. Video source: V5 in Table A.1.

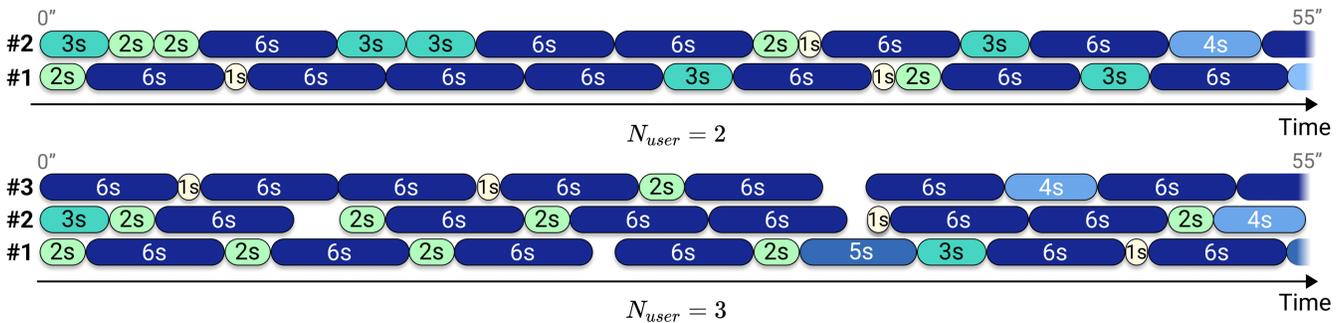

Figure 8: Illustration of comment sequence composition when the maximum number of concurrent comments to be displayed $N_{user}$ is set to 2 and 3. Each colored bar represents a comment from the resulting sequence, its starting position indicating the appearance time and its length indicating the display duration. Video source: V3 in Table A.1.

## 5 Evaluation

We conducted a user study to examine how ComVi enhances viewing engagement compared to existing platforms.

### 5.1 Material

We implemented five conditions (Figure 10): *ComVi*, *YouTube*, *Danmaku*, *YouTube-1ver.*, and *Danmaku-1ver.*. *YouTube* and *Danmaku*

---
[10]Video sources (lengths): V6 (2 minutes 4 seconds), V7 (3 minutes 53 seconds), V8 (6 minutes 19 seconds), V9 (7 minutes 48 seconds), and V10 (10 minutes 12 seconds); see Table A.1.

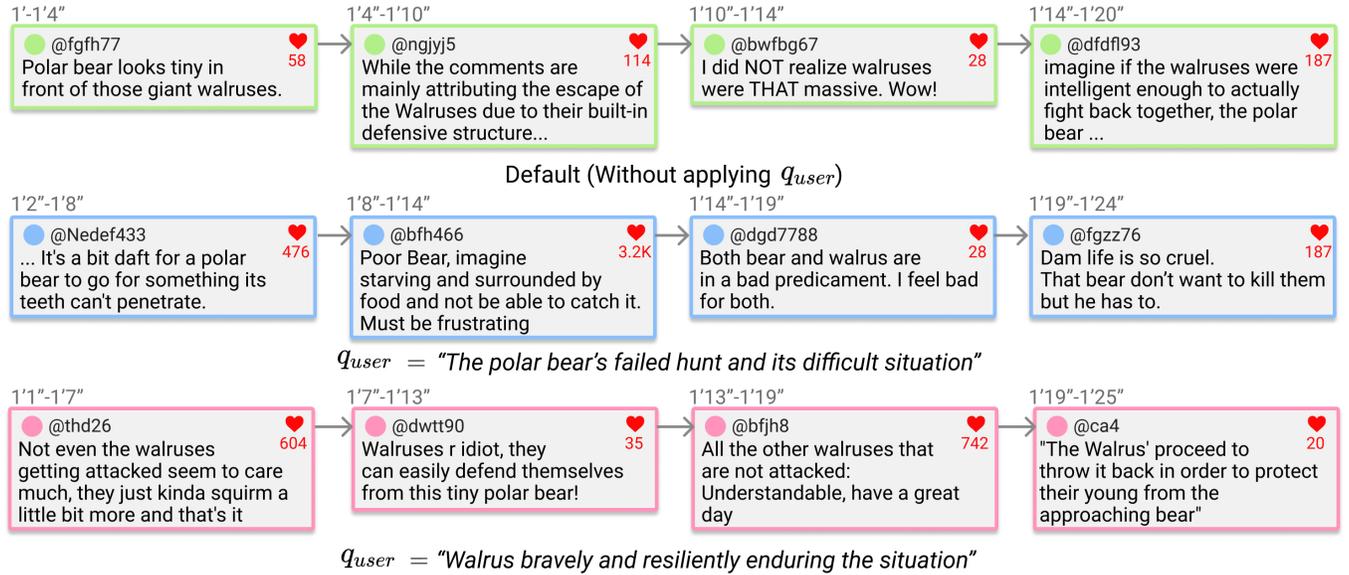

Figure 9: Impact of user-specified query $q_{user}$ on the resulting comment sequence for the same video segment (1 minute to 1 minute 25 seconds). All usernames and profile images were replaced to protect personal information. Video source: V1 in Table A.1.

are implemented to closely resemble their respective platforms, and *YouTube-1ver.* and *Danmaku-1ver.* adapt the display styles of these platforms but show only one comment at a time, similar to *ComVi*'s single-comment format.

For *YouTube*, we emulated its standard layout by placing a scrollable comment list directly below the video player, preserving the original comment order from the actual platform. As general comments do not have a specific timestamp for display, we simulated *Danmaku* by calculating their correlations with all timestamps using Equation (1) and assigning each comment to the timestamp with the highest correlation. To prevent the screen from being fully covered, we set the maximum number of comments displayed at a single timestamp to 20, prioritizing those with the highest correlation scores. During playback, comments appeared at their assigned timestamps, entering from the right boundary of the video frame and traversing horizontally to the left. To reflect the characteristics of the platform where users can set the position and speed of their comments [87], we randomized the vertical placement of each comment and assigned each comment with one of four traverse speeds (140, 180, 220, or 260 pixels per second[11]).

For *YouTube-1ver.*, only one comment is shown at a time, requiring viewers to scroll for every subsequent comment. For *Danmaku-1ver.*, we applied the same correlation-based timestamp assignment as in *Danmaku*, but selected only the comment with the highest correlation score at each timestamp. These selected comments were displayed sequentially, with a new comment appearing only after the previous one had completely exited the screen. We used five videos from diverse genres[12], including movie, entertainment, news, documentary, and music video, and generated the five interface conditions described above.

### 5.2 Procedure

Each participant viewed five interface conditions in a randomized order and watched different videos for each condition to mitigate order and learning effects. Specifically, we prepared ten non-overlapping interface-video pairing configurations and counterbalanced them across participants. The participants were instructed to watch each video with sound enabled and without skipping. After viewing each interface, they were asked to rate the following questions on a 7-point Likert scale:

- **Q1. Mental Demand**: How mentally demanding did you find the experience of reading comments while watching the video? (1: Very low; 7: Very high)
- **Q2. Physical Demand**: How physically demanding did you find the experience of reading comments while watching the video? (1: Very low; 7: Very high)
- **Q3. Contextual Alignment**: How well did the comments align with the video scenes being played? (1: Completely misaligned; 7: Perfectly aligned)
- **Q4. Overall Engagement**: How engaging was the way the comments were presented when you watched the video? (1: Very dissatisfied; 7: Very satisfied)

After completing the evaluation for all five interfaces, participants were asked to select the interface that provided the most satisfying comment-integrated video viewing experience. The study was conducted online and lasted approximately 20 minutes. Participants did not receive monetary compensation.

### 5.3 Results of the User Study

32 participants (15 males, 17 females; age: $M = 26.34, SD = 3.48$) were recruited for the study. Figure 11 presents the results. For the statistical analysis, we conducted Friedman tests on the five interface conditions for each question, with the significance level set at 0.05. The results revealed significant differences among the five interfaces across all questions (Q1: $\chi^2(4) = 76.29$; Q2: $\chi^2(4) = 40.98$; Q3: $\chi^2(4) = 60.1$; Q4: $\chi^2(4) = 46.79$; $p < 0.001$ for all). To further compare *ComVi* with the other four interfaces, we conducted post-hoc pairwise analyses using two-tailed Wilcoxon signed-rank tests with Holm-Bonferroni correction.

---
[11]These speeds were set by measuring comment velocities on *Bilibili*.
[12]See Table A.2 for video information.

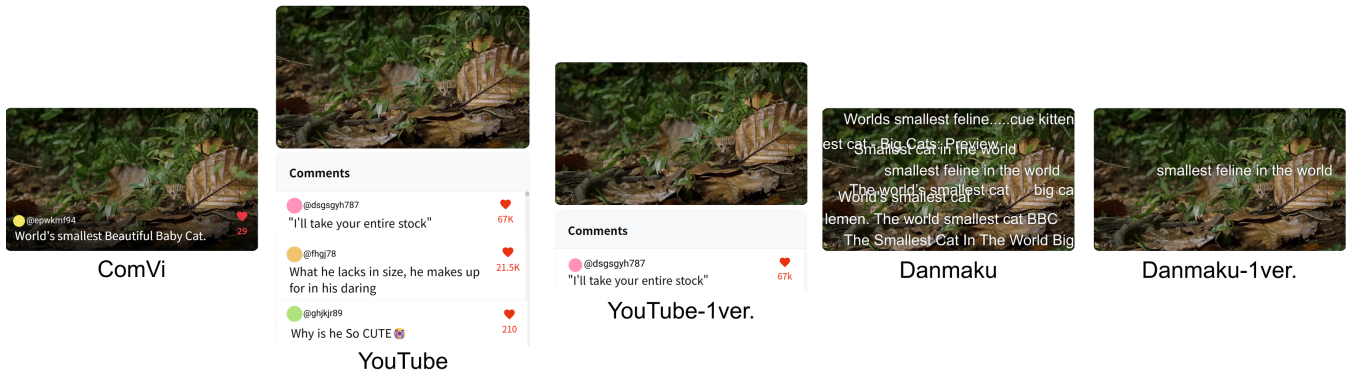

Figure 10: Interface conditions compared in the user study (*ComVi*, *YouTube*, *YouTube-1ver.*, *Danmaku*, and *Danmaku-1ver.*). In the *YouTube* interface, comments are presented in a standard scrollable panel below the video. Following *ComVi* format, each comment includes the user profile and name above the text, with the like count displayed on the right. In the *Danmaku* interface, numerous comments traverse horizontally across the video screen at their assigned timestamps. *YouTube-1ver.* and *Danmaku-1ver.* are single-comment variants of the corresponding interfaces. All usernames and profile images were replaced to protect personal information. Video source: V2 in Table A.1.

In the analysis, *ComVi* ($M = 2.22, SD = 1.36$) was rated significantly less mentally demanding (**Q1**) than *Danmaku* ($M = 6.47, SD = 0.8; p < 0.001$) and *Danmaku-1ver.* ($M = 3.69, SD = 1.89; p < 0.05$). The high mental demand in *Danmaku* can be attributed to the overwhelming volume of comments flowing across the screen. While *Danmaku-1ver.* alleviates this burden through single-comment display, it still imposes greater mental demand than *ComVi*, likely because viewers must continuously chase each moving comment to read it. *ComVi* ($M = 1.5, SD = 0.95$) was rated significantly less physically demanding (**Q2**) than *YouTube* ($M = 3.5, SD = 2.11; p < 0.01$), *YouTube-1ver.* ($M = 3.53, SD = 1.95; p < 0.01$), and *Danmaku* ($M = 4.28, SD = 2.47; p < 0.001$). The high physical demand in *YouTube* and *YouTube-1ver.* can be attributed to manually scrolling and shifting eye gaze between comments and the video, while in *Danmaku*, it results from the rapid eye movements required to track multiple comments moving across the screen.

For contextual alignment (**Q3**), *ComVi* ($M = 5.97, SD = 1.12$) significantly outperformed *YouTube* ($M = 3, SD = 1.41; p < 0.001$), *YouTube-1ver.* ($M = 2.71, SD = 1.65; p < 0.001$), *Danmaku* ($M = 3.75, SD = 1.88; p < 0.001$), and *Danmaku-1ver.* ($M = 5.09, SD = 1.25; p < 0.05$). The poor performance of *YouTube* and *YouTube-1ver.* can be attributed to displaying comments independently of video playback. While *Danmaku* and *Danmaku-1ver.* performed better, they still scored lower than *ComVi* due to the high mental and physical demands that made it difficult for participants to assess contextual alignment.

Finally, *ComVi* ($M = 5.22, SD = 1.72$) achieved significantly higher overall engagement (**Q4**) than *YouTube* ($M = 3.88, SD = 1.62; p < 0.05$), *YouTube-1ver.* ($M = 3.38, SD = 1.64; p < 0.01$), *Danmaku* ($M = 1.72, SD = 1.17; p < 0.001$), and *Danmaku-1ver.* ($M = 3.56, SD = 1.5; p < 0.01$). These results were further reflected in the preference survey, in which 71.9% of participants selected *ComVi* as their most preferred interface, followed by *YouTube* (18.8%), *Danmaku-1ver.* (6.2%), and *YouTube-1ver.* (3.1%), while none chose *Danmaku*.

## 6 Limitations and Future Work

Some viewers may wish to read comments that do not correspond to specific timestamps that express impressions or reactions about the video as a whole (e.g., *"This documentary gave me a completely new perspective!"*). Such overall comments are less likely to be displayed in ComVi because they might exhibit low correlation with specific timestamps. A promising direction for future work is to integrate such overall comments into the ComVi interface. For example, they could be displayed in a separate panel alongside or below the interface, remaining visible regardless of playback.

Comments echoing video content (e.g., quoting narration) are often prioritized due to Sentence-BERT's intrinsic tendency to assign higher similarity scores to lexically overlapping text [75]. While this may help reinforce key content, it could limit opportunities to surface new or diverse perspectives. To balance quoting and novel comments, future direction could add a novelty term $Novel(C_{i,t})$ to the scoring function in Section 3.3, penalizing high lexical overlap [62] or rewarding semantic diversity [20]. Users could also control its strength through a weight parameter.

Diversity within the comment sequence could also be considered to present comments that differ from those already shown. This could be achieved by adding a rule that computes semantic similarity between previously selected comments and the next candidate comment $C_{i',t'}$ (Section 3.2), excluding $C_{i',t'}$ when the similarity is too high (redundancy) or too low (abrupt semantic shifts). To further frame the comment sequence as a story-like progression rather than as isolated utterances, LLMs could assess narrative flow [82], rewarding sequences that follow coherent narrative structures (e.g., a narrative arc) so that they are selected as the optimal sequence. Additionally, textual representations could be extracted at coarser granularities (e.g., actions [86, 89], highlights [28, 80]) beyond the segment-level to capture broader narrative contexts spanning multiple shots or scenes.

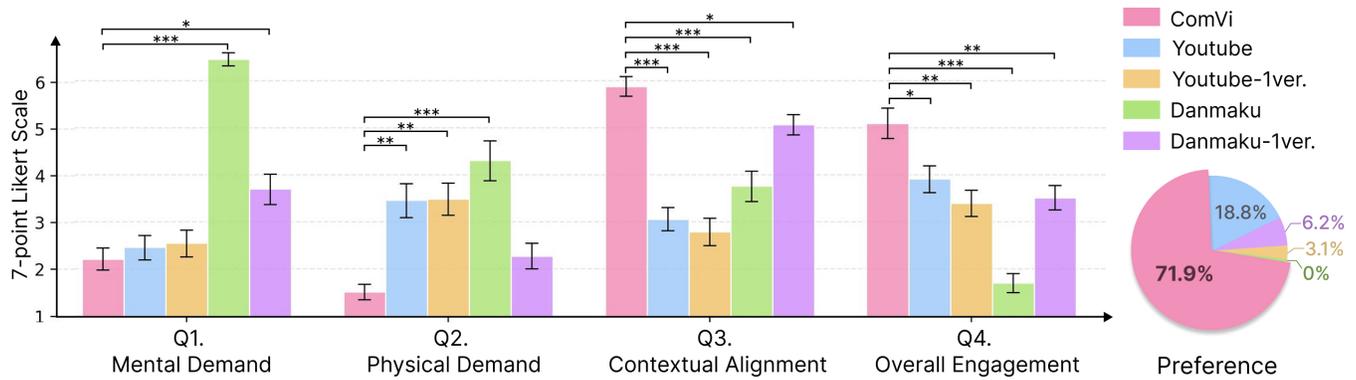

Figure 11: Results from the user study comparing the five interfaces. Error bars represent the standard error of the mean. Asterisks indicate statistically significant differences between pairs of methods (*: $p < 0.05$; **: $p < 0.01$; ***: $p < 0.001$). Significant differences are shown only for comparisons between ComVi and the other four interfaces.

When videos contain high information density or visual complexity (e.g., abundant on-screen text, rapid shot transitions), viewers might experience cognitive overload when simultaneously processing both video content and comments. To mitigate this, $\alpha_{user}$ could be dynamically adapted by detecting such moments [23, 59, 74]. Furthermore, comments could appear more seamlessly by aligning their display timing with natural breaks in the video, such as shot boundaries [36, 49] or speech pauses [35]. ComVi could be extended to support more diverse interactions, such as skipping to the next comment when encountering demanding or uninteresting ones, or continuously displaying additional contextually relevant comments for that timestamp when viewers pause at specific moments.

Currently, ComVi displays comments at the bottom of the screen, potentially obscuring content such as subtitles. Dynamic placement methods [33, 42, 71] can be integrated by positioning comments near key visual elements or placing them to avoid overlapping. Additionally, to enhance readability, excessively long comments could be shortened via summarization [5, 16] or highlighted keywords [37]. Beyond evaluating demand or engagement, future work could examine how viewers divide attention between the video and comments in ComVi—for example, by quantifying gaze distraction through eye-tracking [47] and assessing comprehension of the content [11].

## 7 Conclusion

We propose *ComVi*, a novel system that enables viewers to enjoy comments on general video platforms in a time-synchronized manner by presenting them at semantically relevant playback moments. ComVi first maps each comment to semantically aligned timestamps by computing audio-visual correlations, then selects an optimal comment sequence by balancing temporal relevance, popularity, and adequate display durations. In a user study comparing the comment-reading experience during video playback across ComVi and existing interfaces—YouTube and Danmaku—ComVi provided a significantly more engaging experience, with 71.9% of participants choosing ComVi as their most preferred interface.

## Acknowledgments

This research was supported by Basic Science Research Program through National Research Foundation of Korea (RS-2024-00452748) and by Culture, Sports and Tourism R&D Program through Korea Creative Content Agency (RS-2024-00440434). We thank the anonymous reviewers for their constructive feedback.

# A  Video Materials

Table 1: Data attributes for videos used in Section 4.

| Video ID | Number of Comments | Length | Source |
|---|---|---|---|
| V1 | 14,880 | 3m 34s | *https://www.youtube.com/watch?v=PvWLbK_mNw0* |
| V2 | 84,147 | 1m 57s | *https://www.youtube.com/watch?v=W86cTIoMv2U* |
| V3 | 1,946 | 2m 25s | *https://www.youtube.com/watch?v=S7OWoc-j8qQ* |
| V4 | 80,586 | 2m 36s | *https://www.youtube.com/watch?v=Y2d2HLdBF88* |
| V5 | 7,408 | 4m 52s | *https://www.youtube.com/watch?v=fZLLqapzxRQ* |
| V6 | 1,691 | 2m 4s | *https://www.youtube.com/watch?v=biG3iUCrHaQ* |
| V7 | 1,343 | 3m 53s | *https://www.youtube.com/watch?v=fQUeDdaVoWo* |
| V8 | 1,778 | 6m 19s | *https://www.youtube.com/watch?v=26PrgjTboVQ* |
| V9 | 1,710 | 7m 48s | *https://www.youtube.com/watch?v=SFnMTHhKdkw* |
| V10 | 1,756 | 10m 12s | *https://www.youtube.com/watch?v=6kqaYeY4Sew* |

Table 2: Data attributes for videos used in Section 5.

| Genre | Number of Comments | Length | Source |
|---|---|---|---|
| Movie | 1,946 | 2m 25s | *https://www.youtube.com/watch?v=S7OWoc-j8qQ* |
| Entertainment | 23,068 | 2m 58s | *https://www.youtube.com/watch?v=wFKBN3MGUGI* |
| News | 12,436 | 3m 43s | *https://www.youtube.com/watch?v=RQmqcaS5LIM* |
| Documentary | 84,147 | 1m 57s | *https://www.youtube.com/watch?v=W86cTIoMv2U* |
| Music video | 1,691 | 2m 4s | *https://www.youtube.com/watch?v=biG3iUCrHaQ* |